# A note on transition parameters defined with properties of fluid element

Jianjun Tao

Department of Mechanics and Engineering Science, College of Engineering,

Peking University, Beijing, China

Laminar-turbulent transitions occur at different Reynolds numbers for different flow configurations and different fluids. In order to study quantitatively the similarity among the transition processes of wall-bounded shear flows, a uniform definition of transition parameter is required. The transition parameters defined with properties of fluid element are compared and discussed in terms of theoretical basis, physical definition, and expression, and their application limitations in plane-Poiseuille flow, plane-Couette flow, and Hagen-Poiseuille flow are clarified.

----------------------------

When the external disturbances are strong and effective enough, linearly stable flows may become unstable and the subcritical laminar-turbulent transition occurs, e.g. the boundary-layer flow, the plane-Poiseuille flow (pPF), the plane-Couette flow (pCF), and the pipe flow or Hagen-Poiseuille flow (HPF). In order to describe quantitatively the similarity among the transition scenarios, universal definition of control parameter becomes necessary. Because the fluid elements in different flow configurations are subjected to the same governing equations, one method is to consider the kinematic and dynamic properties of fluid element in the definition.

Defining control or transition parameter with the maximum value of a basic-flow function in the cross section of flow domain may trace back to 1959, when Ryan and Johnson [1] proposed a local parameter $Z_{max}$ ("a function of the ratio of input energy to energy dissipation for an element of fluid") based on the linearized two-dimensional kinetic energy equation to describe uniformly the pipe flow transitions for Newtonian and non-Newtonian fluids,

$$Z = \frac{R\rho U}{\tau_W}\frac{\partial U}{\partial y} . \qquad (1)$$

The subscript "max" represents the maximum value in the cross section of the laminar flow. It is easy to find that $Z_{max} = \frac{2}{3\sqrt{3}}\frac{RU_M}{\nu}$ is obtained at $r=R/\sqrt{3}$ and $U_M$ is the maximum basic-flow velocity. Based on the Navier-Stokes (NS) equation, Hanks extended the Z concept to pPF [2], and named a parameter as K, representing the ratio of energy gradient in the transverse direction to viscous force

($\nabla \cdot \boldsymbol{\tau}$) or pressure gradient in the streamwise direction. For pPF,

$$K = \frac{1}{2}\rho \frac{|\nabla(\boldsymbol{U}\cdot\boldsymbol{U})|}{|\nabla\cdot\boldsymbol{\tau}|} = \frac{1}{2}\rho \frac{d(U^2)/dy}{(\mu\nabla^2 U)_s} = \frac{1}{2}\rho \frac{d(U^2)/dy}{dp/dx}, \tag{2}$$

and K obtains its maximum values $\frac{1}{3\sqrt{3}}\frac{RU_M}{\nu}$ and $\frac{2}{3\sqrt{3}}\frac{hU_M}{\nu}$ at $r=R/\sqrt{3}$ and $y=h/\sqrt{3}$ for pipe flow and pPF, respectively.

Dou [3,4] claims that he proposed a new theory, the energy gradient theory, to describe the turbulent transition, and the central concept of the theory is the energy gradient function K, the ratio of energy gradient in the transverse direction ($\partial E/\partial n$) to the rate of "energy loss" along the streamwise direction ($\partial H/\partial s$).

(a) The question why the flow instability and transition only depend on the ratio between these two terms is mentioned in two aspects: one is "according to observations" [3], and another is a particle model. "If the net energy gained by collisions is larger than this critical amount, this particle will become unstable and move up to neighboring streamline with higher kinetic energy. Similarly, in the second half-period, if the energy released by collision is not zero, this particle will try to move to a streamline of lower kinetic energy." [5] (section 7.1 of [4]). Why a particle "will try" to move to a streamline with lower energy instead of higher energy is not explained. This particle model does not include random process of particle collisions.

(b) For plane-Poiseuille flow (PPF), the energy gradient theory defines [4]

$$K = \frac{1}{2}\rho \frac{d(U^2)/dy}{\partial H/\partial x} = \frac{1}{2}\rho \frac{d(U^2)/dy}{(\mu\nabla^2 \mathbf{U})_s}, \tag{3}$$

the same expression as the K of Hanks (Eq 2). This definition cannot be used for Couette flows, where $(\mu\nabla^2\mathbf{U})_s=0$ for basic-flow states. In order to avoid a zero denominator of Eq. (3), the energy gradient theory modifies the NS equation by adding artificially an additional term W, "work input to per unit volume of fluid by external object/influence" (see Eq. (1) of [6], Eq. (9.1) of [4]), and assumes that W must be considered in Couette flows but not in Poiseuille flows (see Eq. (4) and Eq. (5) in [7], Eq(9.3)and Eq. (9.4) in [4]), ignoring that the fluid elements in Couette flows and Poiseuille flows are governed by the same conservation laws. Consequently, the energy gradient theory defines $\partial H/\partial x$ as $(\tau/u)*(du/dy)$ for pCF and $(\mu\nabla^2\mathbf{U})_s$ for Poiseuille flows [5,6,8] (see also Eq. (9.13) and Eq. (9.14) in [4]), making the energy gradient function K lose the uniformity of definition.

(c) For pCF, the energy gradient theory claims that "Because the energy loss has a damping role to any flow disturbance, the flow near the bottom wall is therefore strongly stable. Towards the top plate, the energy loss is lowest and the flow is therefore most possibly unstable." [6]. Obviously, this conclusion is absurd because the basic flow of pCF is antisymmetric about the midplane, and is caused by the fact that the definition of K does not satisfy the Galilean Invariance.

For the subcritical transitions of viscous shear flows, the nonlinear effect arising from the nonlinear convection term and the viscous diffusion effect brought by the viscous term in the NS equation must be considered. After simplifying the convection and the viscous terms with a perturbation model, a local Reynolds number is proposed by calculating the ratio between these two terms [9], e.g., for channel flows it is

$$\left| \frac{U \frac{\partial U}{\partial y} \eta_0}{\nu \left( \frac{\partial^2 U}{\partial y^2} + \frac{\eta_0}{l} \frac{dU_0}{dy} \right)} \right| \approx Re_m A \qquad (4)$$

where $\eta_0$ and $l$ are local perturbation parameters, $A$ represents the nondimensional perturbation amplitude, and $U_0$ is the basic flow velocity. The local Reynolds number $Re_M$ is the maximum of $Re_m$ in the cross section. For pPF, the relatively smaller term including $\eta_0/l$ in the denominator can be ignored, and it is easy to get $Re_M$ with $A=\eta_0$ by evaluating $U$ with $U_0$. For pCF, $\partial^2 U/\partial y^2$ is nearly zero and the $\eta_0/l$ term becomes dominant. Estimating $U$ with $U_0=U_M(y/h)$ in Eq. (4), it is obtained that $Re_m=|U_0 y/\nu|$ and the perturbation amplitude $A=l/h$, the counterpart of $\eta_0$. Consequently, $Re_M=U_0 h/\nu$ for PCF.

Inspired by the pioneering work of Ryan, Johnson and Hanks[1,2], an energy equation can be obtained by taking the scalar product of momentum equation and the basic-flow velocity [10],

$$\frac{\rho}{2} \frac{d(U_0 U_1)}{dt} = \frac{\rho \nu U_1}{2} \frac{dU_0}{dx_2} + \frac{U_0}{2} \frac{\partial \tau_{1i}}{\partial x_i} \qquad (5)$$

Where $U_1=U_0+u$, $U_2=v$, and $\tau_{1i}$ are the disturbing stress components. The left hand side of Eq. (5) represents the growth rate of total kinetic energy of a disturbed fluid element. On the right side, the first and the second terms denote the rate of energy supplement transferred from the main stream and the energy dissipation rate. Obviously, the right two terms determine the evolution of the kinetic energy, and their ratio is analyzed to derive a uniform definition of local Reynolds number for both Poiseuille flows and Couette flows [10],

$$Re_L = |\rho R_h U \frac{dU}{dx_2}/\tau_w|_{max}, \tag{6}$$

where $R_h$ is the hydraulic radius of the cross section and $\tau_w$ is the wall shear stress.

Table I  Comparison of different local transition parameters

| Author | Theoretical Base | Physical definition | Symbol | Final expression of the parameter | | |
|---|---|---|---|---|---|---|
| | | | | HPF | PPF | PCF |
| Ryan et al. 1959 [1] | Linearized kinetic energy equation | ratio of input energy to energy dissipation for a fluid element | Z | $\frac{2}{3\sqrt{3}}\frac{U_M R}{\nu}$ | | |
| Hanks 1963 [2] | NS equation | ratio of energy gradient in the transverse direction to viscous force or pressure gradient in the streamwise direction | K | $\frac{1}{3\sqrt{3}}\frac{U_M R}{\nu}$ | $\frac{2}{3\sqrt{3}}\frac{U_M h}{\nu}$ | |
| Dou, et al. 2006 [3-8, 11] | Particle model | ratio of energy gradient in the transverse direction to the 'energy loss' in the streamwise direction | K | $\frac{1}{3\sqrt{3}}\frac{U_M R}{\nu}$ | $\frac{2}{3\sqrt{3}}\frac{U_M h}{\nu}$ | X |
| Tao et al. 2011 [9] | NS equation | ratio of the nonlinear momentum advection to the momentum diffusion | $Re_m$ | $\frac{1}{3\sqrt{3}}\frac{U_M R}{\nu}$ | $\frac{2}{3\sqrt{3}}\frac{U_M h}{\nu}$ | $\frac{U_M h}{\nu}$ |
| Tao et al. 2013 [10] | Kinetic energy equation | ratio of input energy to energy dissipation for a fluid element | $Re_L$ | $\frac{1}{3\sqrt{3}}\frac{U_M R}{\nu}$ | $\frac{2}{3\sqrt{3}}\frac{U_M h}{\nu}$ | $\frac{U_M h}{\nu}$ |

HPF: Hagen-Poiseuille flow, PPF: plane-Poiseuille flow, PCF: plane-Couette flow, X: definition changed.

It is shown in Table I, different transition parameters have similar or the same final expressions for given flows, but have different areas of applications. As discussed above, the energy gradient theory uses the same method as Hanks [2] to obtain the transition parameters for Poiseuille flows, and its K has the same expression as the K of Hanks. For Couette flows, however, the energy gradient theory changes K's definition by modifying the NS equation with an additional artificial term, losing the uniformity of definition.

By using the local Reynolds number $Re_L$, the subcritical transition scenarios of pPF, pCF and HPF can be compared quantitatively and illustrate a common sequence of transition stages [10]. The classical Reynolds number considers bulk properties of the flow, while the above discussed local transition parameters embody velocity profile characteristics and may reflect more intrinsic features of the flow field. However, it should be noted that the governing equations are greatly simplified during the derivation process and the final expressions of the local parameters only depend on basic-flow properties and are sensitive to the coordinate settings. Consequently, the local parameter analyses are not expected to be able to explain the rich details during the onset of turbulence. With the development of computational abilities, experimental techniques, dynamic modeling, and stability analysis, it is hopeful to give birth to new transitional parameters to describe more precisely the commonness, similarity, and individuality of subcritical transitions in wall-bounded shear flows.